\documentclass[useAMS,usegraphicx,usenatbib]{mn2e}
\newcommand{\II}{\scriptsize{II}\normalsize}

\usepackage{amsfonts,amssymb,amsmath}
\usepackage{xcolor}
\usepackage{natbib}
\usepackage{hyperref}
\usepackage{graphicx}
\usepackage{soul}
\title[A possible long-term activity cycle for $\iota$ Hor]{A possible
  long-term activity cycle for $\iota$ Horologii: \\First results from
  the HK$\alpha$ \& SPI-HK$\alpha$ projects} 
\author[Flores et al.]{Mat\'ias G. Flores$^{1,2,5}$ \thanks{Visiting
    Astronomer, Complejo Astron\'omico El Leoncito operated under
    agreement between the Consejo Nacional de Investigaciones
    Cient\'ificas y T\'ecnicas de la Rep\'ublica Argentina and the
    National Universities of La Plata, C\'ordoba and San
    Juan.}\thanks{E-mail: 
mflores@icate-conicet.gob.ar}, 
Andrea P. Buccino $^{3,4,5}$, 
Carlos E. Saffe$^{1,2,5}$ and 
 \newauthor 
Pablo J. D. Mauas$^{3,4,5}$\\
$^{1}$ Instituto de Ciencias Astron\'omicas, de la Tierra y del Espacio, Casilla de Correo 49, 5400 San Juan, Argentina\\
$^{2}$ Facultad de Ciencias Exactas, F\'isicas y Naturales, Universidad Nacional de San Juan, San Juan, Argentina.\\
$^{3}$ Instituto de Astronom\'ia y F\'isica del Espacio (IAFE, CONICET-UBA), Casilla de Correo 67, 1428, Buenos Aires, Argentina.\\
$^{4}$ Dpto. de F\'isica, Facultad de Ciencias Exactas y Naturales (FCEN), Universidad de Buenos Aires (UBA), Buenos Aires, Argentina\\
$^{5}$Consejo Nacional de Investigaciones Cient\'ificas y T\'ecnicas (CONICET), Argentina.\\
}

\begin{document}


\pagerange{\pageref{firstpage}--\pageref{lastpage}} \pubyear{2016}

\maketitle

\label{firstpage}

\begin{abstract}

To detect stellar activity cycles and study the possible star-planet
interactions (SPI's), we have developed both  HK$\alpha$ and
SPI-HK$\alpha$ projects since 1999 and 2012 respectively. 

In this work, we present preliminary results of 
possible SPI's studying the chromospheric activity and look for possible correlations 
between stellar activity and stellar/planetary parameters. We find that for stars with similar T$_{eff}$, stellar activity increases with the mass of the planet, similar to previous works. However, stellar ages can also play a role and a larger stellar sample
is needed to verify these trends. We also note that some of 
these stars present a remarkably high level of chromospheric activity,
even comparable with RSCvn or BY Dra active stars. In addition, we
do not observe any correlation between stellar activity and semi-major
axis. 

We present the first long-term activity study of the star $\iota$
Horologii, a young solar-type star which hosts a non-transiting Jovian
planet and presents a high activity level. We analyze our own spectra,
obtained between 2002 and 2015, combined with public HARPS
observations. We calculate the Ca {\sc ii} 
indexes derived from the 987 CASLEO and HARPS spectra and convert
them to the Mount-Wilson scale. We found a long-term activity cycle of
$\sim$ 5 years which fits the \emph{active} sequence of
Bohm-Vitense. The amplitude of this longer cycle is
irregular, as was also observed for the shorter one. This fact could be
attributed to an antisymmetric distribution of active regions on the
stellar surface.
\end{abstract}

\begin{keywords}
 stars: activity-- planet-star interactions -- stars: $\iota$ Horologii.
\end{keywords}

\section{Introduction}

In 1966, Olin Wilson  initiated  the most-extended program dedicated
to the study of the chromospheric activity of several late-F to early-M stars
by monitoring the cores of the H and K Ca \II\- lines at 3968 and 3934
\AA \ \citep[see details in][]{1995ApJ...438..269B}.
This program, performed at Mt. Wilson Observatory between 1966 and 2003,
together with other similar studies,  allowed to detect
chromospheric activity cycles in  several stars of different spectral types,
with periods  ranging between 2.5 and 25 years
\citep[e.g.][]{1978ApJ...226..379W, 1995ApJ...438..269B,
  2008A&A...483..903B, 2016A&A...589A.135F}. The 
usually accepted model to interpret these stellar cycles is the
$\alpha\Omega$-dynamo, which proposes that the generation and
intensification of magnetic fields in these stars is produced by a
global-scale dynamo action arising from the coupling of convection and
rotation \citep{1955ApJ...122..293P,1982AA...108..322}.

For a sample of stars with well-determined rotation and activity-cycle
periods, \cite{1999ApJ...524..295S} examined the relation between both
frequencies and found that most cycles fall into two branches, which
they classified as the \emph{active} and \emph{inactive} sequences. In
particular, a few stars exhibit cycle periods on both branches
simultaneously. For instance, the star HD 190406 (G0V) shows
signatures of a short secondary cycle of 2.6 years superimposed on its
much longer primary cycle of 16.9 years
\citep{2007AJ....133..862H}. Other stars with multiple
  activity cycles were reported by \citet{2009A&A...501..703O}, who
  studied the time variations of 20 cyclic active stars from
  decade-long photometric and spectroscopic observations. As a result,
  they found that 75\% of them present multiple cycles with variable
  cycle lengths. More recently, \citet{2013ApJ...763L..26M} detected
  two simultaneous cycles with periods of 2.95 and 12.7 years in the
  K2V star $\varepsilon$ Eridani (HD 22049). To interpret the
  $P_{cyc}-P_{rot}$ bimodal distribution, \cite{2007ApJ...657..486B}
suggests that probably two different dynamos are operating inside
these stars: one driven by rotational shear in the near-surface layers
responsible for the longer cycle (active branch), and the other one
driven by a so-called tachocline at the base of the outer convection
zone (shorter cycle).   

 $\iota$ Horologii (HD17051 = HR 810, $ V = 5.4$, $B-V = 0.57$)
is a young solar-like F8V star.  \cite{2000A&A...353L..33K} detected a
non-transiting Jovian planet (2.26 $M_{Jup}$) at a distance $a=0.925$
AU orbiting $\iota$ Hor with an orbital period of 320.1 days. Although
it is a southern star, \cite{2008A&A...482L...5V} found evidences in
the acoustic oscillation frequencies of the star that $\iota$ Hor had
evaporated from the primordial Hyades cluster, as
\cite{2001MNRAS.328...45M} have previously deduced from a kinematic
analysis.  In particular, \cite{2008A&A...482L...5V} estimated the age
of $\iota$ Hor as 625 Myr.  On the other hand,
\cite{2010ApJ...723L.213M} estimated a rotation period between 7.9 and
8.5 days for $\iota$ Hor, consistent with the mean rotation period of
solar-type stars in the Hyades cluster \citep{1995ApJ...452..332R}. Due
to its short rotation period and young age it is expected that $\iota$
Hor presents a higher level of activity than the Sun, in
agreement with the widely believed close connection between rotation
and activity \citep{1984ApJ...279..763N}. 

From Ca \II\- H and K observations obtained between 2008 and 2010,
\cite{2010ApJ...723L.213M} reported a 1.6-yr chromospheric cycle for
this star, which is one of the shortest chromospheric cycles reported
for solar-type stars\footnote{Recently, using spectropolarimetric observations, \citet{2016MNRAS.459.4325M} detected a chromospheric activity cycle ($\sim$ 0.32-yr) in the exoplanet host star $\tau$ Bo\"{o}tis (F7V), being the shortest detected cycle up to now.}. Following the reasoning of
 \cite{2007ApJ...657..486B}, they 
 associated this short cycle to the \emph{inactive} branch of the $P_{cyc}-P_{rot}$
 diagram and suggested that the cycle corresponding
 to the \emph{active} branch would be close to 6
 years. \cite{2013A&A...553L...6S} also detected this 1.6-yr cycle in
 the X-ray emission, which is the first coronal cycle reported in a
 solar-like star. They also propose that this short coronal cycle is
 likely modulated by a longer one. However, this possible 
 longer activity cycle has not been reported in the literature to date. 
  
The discovery of activity cycles in stars with physical
characteristics similar to the Sun across a range of age and other
parameters is important to understand how typical the Sun is. Also,
understanding stellar activity is necessary to discriminate between stellar activity and
planetary signals in radial-velocity surveys
\citep[e.g.][]{2013ApJ...774..147R,2014A&A...567A..48C}. For instance,
\citet{2011A&A...528A...4B} separated stellar and planetary signals
in radial velocity measurements of $\iota$ Hor. To do so, they 	 
carried out a simultaneous modeling of stellar activity and planetary
parameter for this star. As a result, the authors excluded the
presence of a second planet in this system. This study reveals the
importance of monitoring the chromospheric stellar activity of stars.

To study the long-term activity in solar-type stars, in 1999 we have
developed the HK$\alpha $ project, an observing program at the Argentinian observatory Complejo
Astron\' omico El Leoncito (CASLEO), dedicated to periodically obtain 
medium resolution echelle spectra of a sample of F to M southern stars \citep[see details in][]{2011AJ....141...34B}. Since then, we have systematically observed more
than 150 main-sequence stars from F5.5 to M5. To date, we have more
than 5500 mid-resolution echelle spectra (R $\sim$ 13000) ranging
from 3890 to 6690 \AA \ \citep{2005A&A...441..701V, 2007MNRAS.378.1007D,
2008A&A...483..903B, 2009MNRAS.398.1495V, 2013ApJ...763L..26M}.
Based on our long spectra database, we found the first chromospheric
activity cycles in M stars
\citep{2007A&A...461.1107C, 2007A&A...474..345D, 2011AJ....141...34B, 2014ApJ...781L...9B}. 

On the other hand, the detection of extrasolar planets around
solar-type stars opened a new discussion related to the possible influence of the planet
on the level of activity of its host star. On average, 7$\%$ of these
exoplanets confirmed to date are hot Jupiters (orbital period less
than 7 days, a semi-major axis less than 0.1 AU, and a minimum mass
greater than 0.2 times the mass of Jupiter) in orbits around F to K
solar-type stars\footnote{http://exoplanets.eu}. Over this last decade, several works
\citep{2008ApJ...676..628S, 2008A&A...482..691W, 2008MNRAS.385.1179D,
  2012A&A...540A..82K, 2014A&A...565L...1P}
showed that the level of activity of the host star can be influenced
by its interaction with the exoplanet. In order to explore the
star-planet interactions (SPI's) in southern stars, 
in 2012 we started the SPI-HK$\alpha $ project, a branch of the
HK$\alpha $ program, focused specially on stars with hot
Jupiters or massive planets in eccentric orbits, which are the
best candidates to study SPI's. To compare the
level of activity of the exoplanet host stars, we  have also taken
spectra of stars without planets of similar spectral types.  

In particular, we have been monitoring the activity of $\iota$ Hor
since 2002 and, as a result, we have a long time-series of Mount
Wilson indexes obtained between 2002 and 2015, which is an useful tool
to search for the possible longer activity cycle suggested by
\cite{2010ApJ...723L.213M} and \cite{2013A&A...553L...6S}. In this
work we present a complete analysis of this data, organized as
follows. In Section \S\ref{sec.obs} we give an overview of our data,
which is complemented by data obtained with HARPS and by the SMARTS survey. In Section
\S\ref{sec.res} we describe our results, and finally we outline our
discussion in Section \S\ref{sec.disc}.  

\section{Observations}\label{sec.obs}
 
\noindent The observations were obtained at the Argentinean
Observatory Complejo Astron\'omico El Leoncito (CASLEO, San Juan) with
the REOSC\footnote{http://www.casleo.gov.ar/instrumental/js-reosc.php}
spectrograph located in the 2.15 m telescope. The CASLEO 
spectra were reduced by using standard IRAF\footnote{\textsf{IRAF is
    distributed by the National Optical Astronomy Observatory, which
    is operated by the Association of Universities for Research in
    Astronomy, Inc. under cooperative agreement with the National
    Science Foundation.}} routines, which carry out the usual
reduction process including bias subtraction, spectral order
extractions, wavelength calibration and other operations. All the
spectra from the HK$\alpha$ and SPI-HK$\alpha$ projects are
flux-calibrated. To do so, for each star we also acquired a long-slit
spectrum, which was flux-calibrated by using a set of
spectrophotometric standards stars. This spectrum was then used to
obtain a sensitivity function. As a final step, we combined the
flux-calibrated echelle orders to obtain a one-dimensional spectra
\citep[see][for details]{2004A&A...414..699C}.

In Table \ref{tab.one} we show the Mount Wilson $S${\tiny
    $\mathsf{_{MW}}$} index, which essentially is the
ratio of the flux in the core of the Ca \II\- H and K lines to the
nearby continuum 
\citep{1978PASP...90..267V}. To compute $S${\tiny
    $\mathsf{_{MW}}$}, for each spectrum we
integrated the flux in two windows centered at the cores of the Ca
\II \ lines, weighted with triangular profiles of 1.09 \AA \ FWHM, and computed the ratio of these fluxes to the mean continuum
flux, integrated in two passbands of 20\AA\--wide centered at
3891\AA\- and 4001\AA. Finally, we used the calibration between Ca \II\-
indexes derived from the CASLEO spectra and the Mount Wilson indexes
obtained by \citet{2007A&A...469..309C}.\\



\begin{table}
\centering
\caption{Observational data and derived Mount Wilson activity indexes
  for the star $\iota$ Hor from the HK$\alpha$ \& SPI-HK$\alpha$
  projects.} 
\label{tab.one}
\begin{tabular}{|c|c|c|c|} 
\hline
\hline
Date$^{*}$ & JD (+2,450,000) &  $S${\tiny $\mathsf{_{MW}}$} & $\sigma_{S}$\\
\hline
  0603 & 2803 & 0.213 & 0.0070\\
  0607 & 4282 & 0.232 & 0.0078\\
  0608 & 4640 & 0.243 & 0.0083\\
  0706 & 3926 & 0.231 & 0.0078\\
  0710 & 5402 & 0.237 & 0.0080\\
  0802 & 2517 & 0.201 & 0.0066\\
  0805 & 3597 & 0.250 & 0.0086\\
  0813 & 6537 & 0.255 & 0.0088\\
  0903 & 2895 & 0.222 & 0.0074\\
  0904 & 3274 & 0.247 & 0.0085\\
  0905 & 3636 & 0.246 & 0.0084\\
  0906 & 3989 & 0.224 & 0.0075\\
  0908 & 4732 & 0.222 & 0.0073\\
  0910 & 5455 & 0.246 & 0.0084\\
  0911 & 5820 & 0.240 & 0.0082\\
  0912 & 6171 & 0.228 & 0.0076\\
  0912 & 6196 & 0.224 & 0.0075\\
  0913 & 6564 & 0.257 & 0.0089\\
  0914 & 6905 & 0.276 & 0.0097\\
  1214 & 7004 & 0.265 & 0.0093\\
  1009 & 5107 & 0.242 & 0.0083\\
  1013 & 6592 & 0.253 & 0.0087\\
  1102 & 2602 & 0.195 & 0.0063\\
  1112 & 6254 & 0.244 & 0.0083\\
  1203 & 2979 & 0.225 & 0.0075\\
  1207 & 4447 & 0.230 & 0.0077\\
  1208 & 4821 & 0.234 & 0.0079\\
  1210 & 5547 & 0.248 & 0.0085\\
  1212 & 6283 & 0.225 & 0.0075\\
  0915 & 7273 & 0.223 & 0.0074\\
\hline
\multicolumn{4}{c}{Note: $^{*}$ Month and year of the observation}\\ 
\multicolumn{4}{c}{(e.g., 1102 correspond to 2002 November).}\\
\end{tabular}
\end{table}



\noindent We complement our analysis with public high-resolution
spectra (R $\sim$ 115.000) taken between 2003 and 2015 with the HARPS
(High Accuracy Radial velocity Planet Searcher) spectrograph,
installed at the 3.6 m ESO telescope under the programs listed in Table
  \ref{tab.two}. These spectra cover the spectral range
3781-6912~\AA~and were automatically processed by the HARPS
pipeline\footnote{http://www.eso.org/sci/facilities/lasilla/instruments/harps/}. We
carefully selected only those spectra with high signal-to-noise ratio
to measure precisely the Ca \II\- H and K line-core fluxes. Then,
we derived the Mount Wilson S indexes for 957 spectra with a mean S/N
$\sim$ 175 at 6490~\AA. To do so, we used the calibration procedure
explained in \citet{2011arXiv1107.5325L}. 

In addition, we also included the Mount Wilson indexes 
obtained between 2008 and 2010 by \citet{2010ApJ...723L.213M}, and those
acquired between 2010 and 2013 presented by \citet{2013A&A...553L...6S}. 
These indexes were derived from spectra taken with RC spectrograph (R $\sim$ 2500) on
the SMARTS (Small and Moderate Aperture Research Telescope System) 1.5
m telescope at Cerro Tololo Interamerican Observatory (CTIO), and
reduced with a standard procedure \citep[see details in][]{2010ApJ...723L.213M}.

\begin{table}
\centering
\caption{ID programs of $\iota$ Hor observations used in this work.}
\label{tab.two}
\begin{tabular}{|c|}  
\hline
\hline
\multicolumn{1}{|c|}{ESO HARPS programs}\\
 \hline
 078.D-0067(A)\\ 
 077.C-0530(A)\\ 
 079.C-0681(A)\\ 
 078.C-0833(A)\\
 060.A-9036(A)\\
 072.C-0531(B)\\
 072.C-0488(E)\\
 074.C-0012(A)\\
 073.C-0784(B)\\
 076.C-0878(A)\\
 091.C-0853(A)\\ 
 060.A-9700(G)\\
\hline
\end{tabular}
\end{table}
  
\section{Results}\label{sec.res}
\subsection{The first SPI-HK$\alpha$ project results}

\begin{figure}
 \includegraphics[scale=0.14]{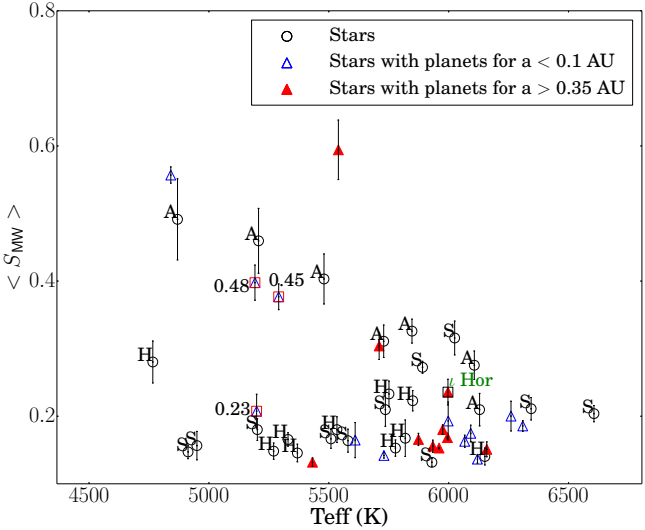}
 \caption{Mean activity level $<S${\tiny $\mathsf{_{MW}}$}$>$
     vs. T$_{eff}$ for stars with and without planets. Empty blue
     triangles are the stars with hot Jupiters planets (a $\leq$ 0.1
     AU). Stars with planets located at ``larger'' orbital separation
     are indicated with full red triangles (a $>$ 0.35 AU). Empty
     black circles correspond to stars without planets (Single,
     labeled with S), very active BY Dra and RSCvn stars (labeled with
     ``A'') and high-proper motion stars (labeled with ``H''). Empty blue triangles inside red
     squares indicate those stars of similar T$_{eff}$ (T$_{eff}$ $\leq$ 5500 K) with hot Jupiters planets which show high activity levels (labeled with their planets masses). The star $\iota$ Hor is indicated with a full red triangle inside a black square.} 
\label{fig.teff}
\end{figure}

In Fig.~\ref{fig.teff} we present
the mean Mount Wilson index $<S${\tiny $\mathsf{_{MW}}$}$>$ distribution of
exoplanet host stars under study as a function of the effective
temperature. All stellar and planetary parameters (e.g. effective
temperature T$_{eff}$, semi-major axis $a$, etc.) were obtained from
\emph{The Extrasolar Planets 
  Encyclopaedia }\footnote{http://exoplanets.eu}. For
comparative purposes, we also include in the figure both the mean Mount
Wilson indexes of single stars (without planets) of similar spectral
type and those corresponding to very active stars as RSCvn and BY Dra type
ones. The index $<S${\tiny $\mathsf{_{MW}}$}$>$ was calculated from
the CASLEO spectra. In 
Fig. \ref{fig.teff}, we note that 
those stars with T$_{eff}$ $\leq$ 5500 K which host close-in 
(a $\leq$ 0.10 AU) and more massive exoplanets show elevated activity levels (empty blue triangles inside red squares), similar to that suggested by \cite{2012A&A...540A..82K}. However, this result could be related to the widely believed close connection between rotation (age) and activity. On the other hand, the activity level of hotter stars (T$_{eff}$ $\geq$ 5500 K) with
  planets are in an intermediate activity zone, between very active as
  RSCvn or BY Dra-type stars and single stars of similar spectral
  type. In particular, $\iota$ Hor 
(full red triangle inside a black square in Fig.\ref{fig.teff}) is one of the most active
  exoplanet host star of our sample. In section
  \S\ref{subsec.longiota} we analyze in detail its long-term
  activity.

\begin{figure}
 \includegraphics[scale=0.14]{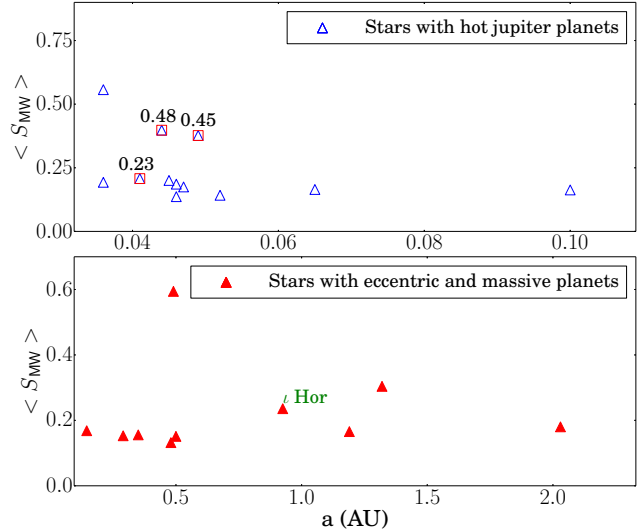}
 \caption{Mean activity level vs. semi-major axis for stars with
planets. Top panel: Empty blue triangles correspond to stars with hot
Jupiters planets. Empty blue triangles inside red squares and labels indicate the position and the
planet masses of the stars with similar T$_{eff}$ plotted in figure
\ref{fig.teff}. Lower panel: Stars which host massive planets in
eccentric orbits are shown with full red triangles. We also indicate
the position of the star $\iota$ Hor.} 
 \label{fig.ahotjup}
\end{figure}

\cite{2000ApJ...533L.151C} suggested that planets located at distances
$\leq$ 0.1 AU could affect their host stars by magnetic and/or tidal
SPI, increasing its chromospheric activity. To search for
observational evidences of this increase, in Fig.~\ref{fig.ahotjup}
we plot $<S${\tiny $\mathsf{_{MW}}$}$>$ vs. semi-major axis $a$ for a
set of 12 stars with hot Jupiters observed at CASLEO. It can be seen
that there is no significant correlation between stellar activity and
semi-major axis. These are, however, preliminary results that should be
confirmed with future observations. 

\subsection{Long-term activity of $\iota$ Hor}\label{subsec.longiota}

\begin{figure}
  \centering
 \includegraphics[scale=0.14]{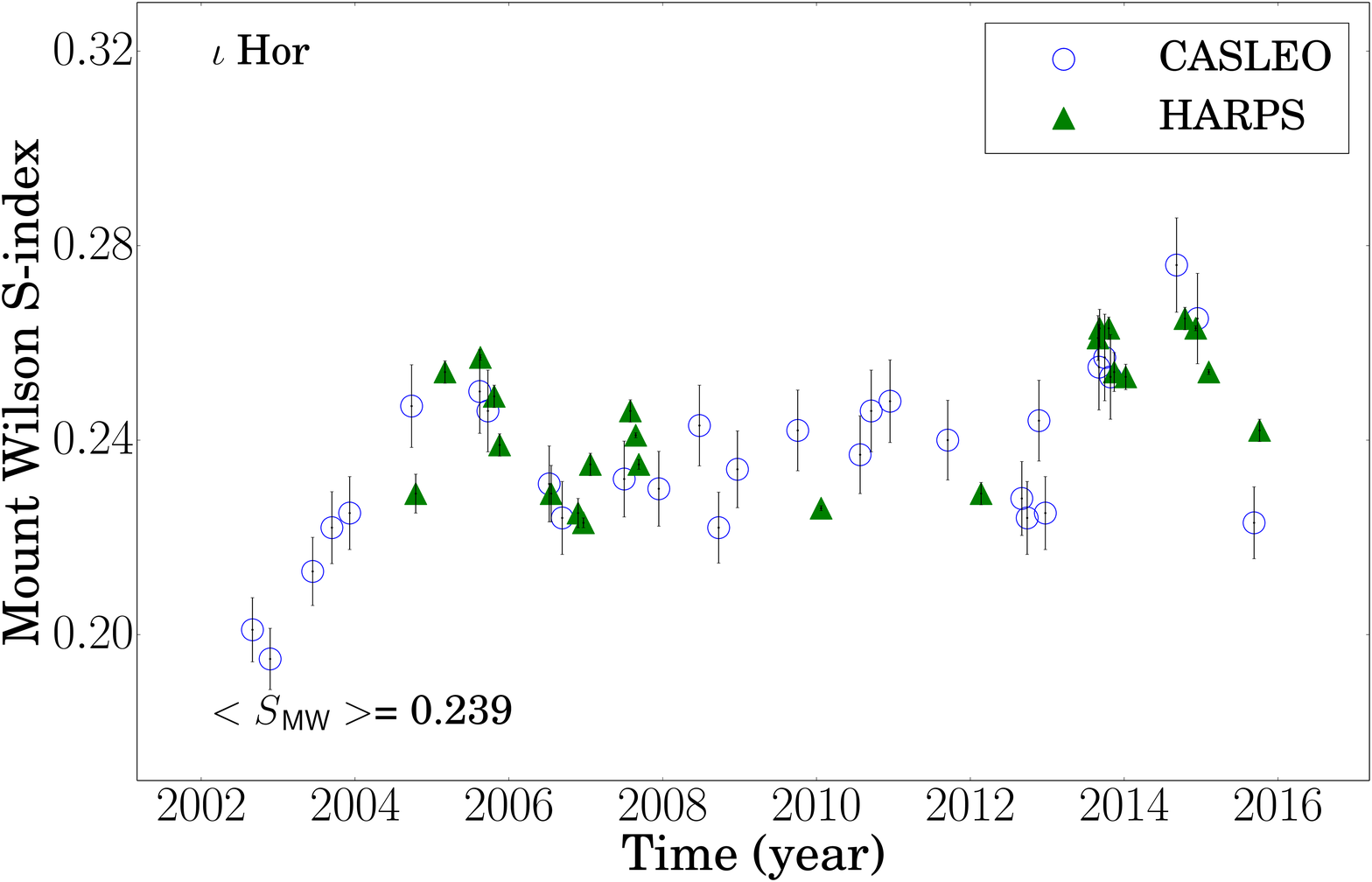}
 \caption{$\iota$ Hor. Time series of Mount Wilson index from combined
   observations. CASLEO data set are shown with empty blue circles and
   HARPS monthly means with full green triangles.}\label{fig.timea} 
\end{figure}

In Fig.~\ref{fig.timea} we show the entire data set collected between
2002 and 2015 for $\iota$ Hor at CASLEO, combined with the Mount Wilson
indexes derived from HARPS spectra obtained between 2003 and 2015. We
resampled the HARPS data by using monthly means.  

We estimated the uncertainty of each CASLEO index as the
nightly variation during an observing run. To do so, we observed
$\iota$ Hor during 5 consecutive nights in November and December
2012. We obtained that the Mount Wilson indexes varies between 1\% to
3\%. Therefore, we considered a
conservative uncertainty of 3\% for each single
observation. On the other hand, we included the average values
of HARPS observations associated to the same observing season (HARPS
monthly means). Error bars of the average values correspond to the
standard deviation of the mean. For those cases with only one observation
in a month, we adopted the typical RMS dispersion of other bins.
Fig.~\ref{fig.timea}, shows a clear agreement between both time series
(CASLEO and HARPS). 

\begin{figure}
  \includegraphics[width=8.5cm,height=6cm]{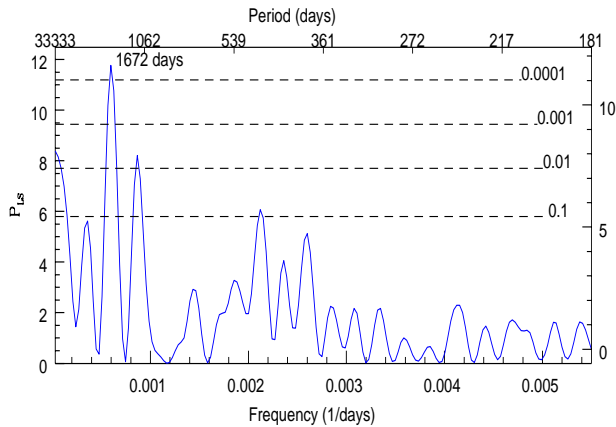}
 \caption{Lomb-Scargle periodogram of the $S${\tiny
       $\mathsf{_{MW}}$} index from the combined data plotted in
     Fig. \ref{fig.timea}.} 
 \label{fig.periodoa}
\end{figure}

To study the long-term chromospheric activity for this combined
datasets we computed the Lomb-Scargle periodogram
\citep{1986ApJ...302..757H} as well as the False Alarm Probability
(FAP) of the periods, which were obtained by using a Monte Carlo
simulation, as described in \cite{2009A&A...495..287B}. We observed a
long period modulation of the activity with a cycle of 1672 $\pm$ 51
days ($\sim$ 4.57 years) with a Monte Carlo FAP $= 2 \times 10^{-5}$
(Fig.~\ref{fig.periodoa}). However, we do not detect a significant
peak near 1.6 years, which is  the one detected by 
\cite{2010ApJ...723L.213M}. 

According to \cite{2007ApJ...657..486B}, in stars with deep convective
zones two active dynamos could be operating simultaneously in
different regions of the star and, as a consequence, give rise to
two different chromospheric activity cycles. Considering the 8.1-day
rotation period of $\iota$ Hor, if the 1.6-yr cycle detected by
\cite{2010ApJ...723L.213M} corresponds to the \emph{inactive} branch,
the 4.57-yr period detected in
this work could correspond to the \emph{active} branch, within the
statistical dispersion of the data plotted in Fig. 1 of
\cite{2007ApJ...657..486B}.

\begin{figure}
  
 \includegraphics[width=8.5cm,height=6cm]{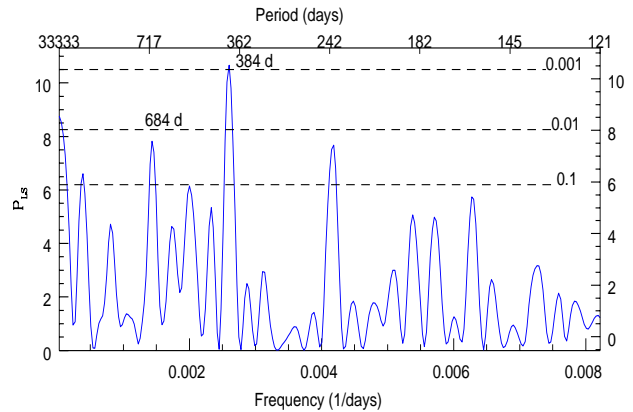}
 \caption{Lomb-Scargle periodogram of the HARPS and CASLEO spectra
   after removing the  harmonic curve of 1672-day period.} 
 \label{fig.periodob}
\end{figure}

To look for short-term variations, we studied the HARPS and CASLEO indexes
after removing the harmonic curve of 1672-day period (see
Fig. \ref{fig.periodob}). We obtained a significant period $P= (384\pm 2)$ 
days with FAP=0.0007. However, we suspect that the 384-day period is an 
artifact. To explore this hypothesis, we used the daily sunspot
numbers taken from the National Geophysical Data
Center\footnote{https://www.ngdc.noaa.gov/} between 1700 
and 2016. We took a sample of the solar data with the same phase
intervals that our data, we added Gaussian
errors of 10\% at each point, and we computed the
Lomb-Scargle periodogram. To take different phases of the solar cycle,
we repeated this proccedure 1000 times with different starting
dates. For each periodogram, we considered the period 
with maximum significance (or minimum FAP) as the detected period. We
obtained 72\% of the detected period between 9.12 and 13.03 years with
FAP$<$0.001, which corresponds to the solar-cycle, and in 670  of
these 1000 periodograms significant periods with FAP$<$0.10  between
326 and 441 days (384 day $\pm$15\%) are also present.  This results
support the fact that the 384-day peak detected is an artifact. 

In  Fig. \ref{fig.periodob}, we also detected a less prominent peak
at ($684\pm 8$) days (1.87 years) with FAP=0.016, which could be
associated to the short-term cycle detected by 
\cite{2010ApJ...723L.213M}.

On the other hand, in Fig. \ref{fig.timea} it can be observed that
between 2007 and 2012 the $S${\tiny $\mathsf{_{MW}}$} index is rather
constant. The Mount Wilson index during this
interval varies slightly with an amplitude $\Delta S=0.013$,
from $S${\tiny $\mathsf{_{MW}}$}$=0.248$ to 
$S${\tiny $\mathsf{_{MW}}$}$=0.222$.
On the other hand, between 2002 and 2007 the amplitude of the
variation is $\Delta S=0.031$, with a maximum of
$S${\tiny $\mathsf{_{MW}}$}$=0.257$ and a minimum of $S${\tiny
$\mathsf{_{MW}}$}$=0.195$, and between 2012 and 2015 the amplitude is 
$\Delta S=0.026$, ranging from $S${\tiny $\mathsf{_{MW}}$}$=0.224$ to
$S${\tiny $\mathsf{_{MW}}$}$=0.276$.

It is broadly known that from 1645 to 1715 solar activity was flat and 
the number of sunspots was extremely reduced, although they did not
disappear. During this period, called Maunder Minimum, it is estimated
that the Ca \II\- K emission was 11\% lower than during the current Solar Minimum
\citep{1992PASP..104.1139W}.  From the analysis of Mount Wilson
indexes registered over 30 years, \cite{1995ApJ...450..896B} reported
that 15\% of the 111 stars under study are ``flat stars''. Some of
them are known to be evolved from the main sequence
\citep{2004AJ....128.1273W}, while others are dwarf stars that could be in a
specially inactive phase like the Maunder Minimum \citep{2007ApJ...663..643J}.  In particular,
the active young solar-like star $\iota$ Hor could belong to this second group.

Unlike the Solar Maunder Minimum, we observe that the level
of activity of $\iota$ Hor during the ``flat'' interval is larger
($<S${\tiny $\mathsf{_{MW}}$}$>$ $\sim 0.237$) than the minima of its
cyclic phase,  
reached in 2002 ($S${\tiny $\mathsf{_{MW}}$}$= 0.195$) and in 2015
($S${\tiny $\mathsf{_{MW}}$}$=0.224$). Similarly, 
$\psi$ Ser (HD 140538) also showed a transition from a constant period
to a cycling mode, and also in this case the Ca \II\- fluxes in the cycle
minima were slightly below that observed during the flat phase 
\citep{2007AJ....133..862H}.

On the other hand, between 2008 and 2010 $\iota$ Hor was continuously
monitored by the SMARTS Southern HK project
\citep{2009arXiv0909.5464M}, which is a time-domain survey of stellar
activity variations. In contrast with our results,
\cite{2010ApJ...723L.213M} detected the 1.6-year cycle during the
``flat'' interval observed in our Fig. \ref{fig.timea}. They
continued observing this star until 2013 and confirmed this
1.6-year cycle, which was also observed in X-rays
\citep{2013A&A...553L...6S}. However, they found that the amplitude of the three 1.6-year cycles observed between 2008 and
2012 decreased from $A_{cyc}\sim 0.024$ to $A_{cyc}\sim 0.012$
(see Fig. 1 in \citealt{2013A&A...553L...6S}).

\begin{figure}
  \centering
 \includegraphics[scale=0.14]{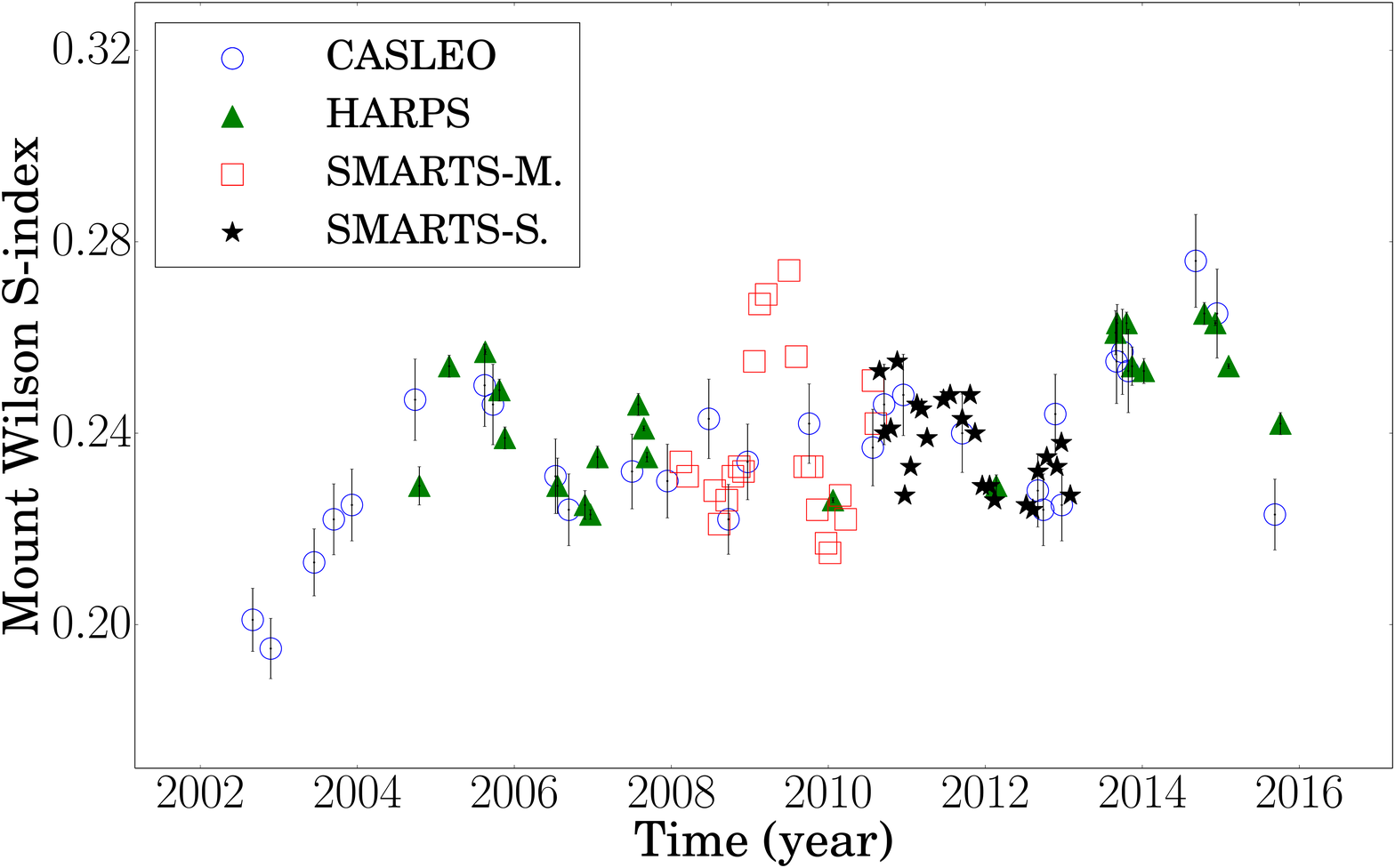}
 \caption{$\iota$ Hor. Time series of chromospheric activity
     indexes from combined observations. CASLEO data are shown
     with empty blue circles and the HARPS monthly means with full green
     triangles. SMARTS data from \citet{2010ApJ...723L.213M} and from 
     \citet{2013A&A...553L...6S} are shown with empty red squares
     and black stars, respectively.}\label{fig.timeb} 
\end{figure}

In figure \ref{fig.timeb} we show our data combined with the
average indexes obtained by SMARTS \citep{2010ApJ...723L.213M, 2013A&A...553L...6S}. We performed a period analysis 
for all the combined data set. As a result, we obtained a Monte Carlo FAP of $= 2 \times 10^{-5}$ for our $\sim$ 5-yr cycle and       
 a FAP of $= 3.61 \times 10^{-4}$ for the shorter one (1.6-yr), much more significant than the one detected in Fig. \ref{fig.periodob}. In addition, it can be observed that between 2007 and 2010 (first 1.6-yr cycle)
the SMARTS indexes vary with an amplitude comparable to those observed
in CASLEO during the intervals 2002-2007 and 2012-2015 ($A_{cyc}\sim
0.029$ vs. $A_{cyc}\sim 0.031$ and $A_{cyc}\sim 0.026$. On the other
hand, the cycle around 2012 has a much smaller amplitude ($A_{cyc}\sim
0.007$).        
Therefore, the ``flat'' interval observed in Fig. \ref{fig.timea} is
not a constant phase, but a short activity cycle with a declining amplitude,
as suggested by \citet{2013A&A...553L...6S}. In other words, the combined time-series suggest that the flat period seen in the CASLEO data is only caused by sampling.

\section{Discussion}
\label{sec.disc}

\citet{2012A&A...540A..82K} found evidence that, for stars
with T$_{eff}<$ 5500 K hosting planets closer than 0.15 AU, their  
stellar activity is correlated with the mass of the planet. To check
this result, we studied 12 stars with hot
Jupiters. We observe that for a similar T$_{eff}$, stellar activity
increases with the mass of the planet. Nevertheless, this trend could be related to differences on the stellar age. To explore this possibility, we used the stellar ages computed by \citet{2016A&A...585A...5B}, which are 11.9, 5.9 and 5.1 Gyr for
less to more active stars,  respectively (corresponding to the empty blue triangles inside red squares of figure \ref{fig.teff}). These data suggest that the observed stellar activity increases  are probably the product of differences in stellar ages and are not related to the presence of massive planets. However, recently \citet{2013A&A...551L...8P} showed that the reduction in the chromospheric activity with age is limited to stars younger than about 2.0 Gyr. On the other hand, we show the distribution of $<S${\tiny$\mathsf{_{MW}}$}$>$ as a function of semi-major axis. We do not observe any correlation between stellar activity and semi-major axis. A larger sample of stars with and without planets is needed to confirm these preliminary results.

We also studied the long-term activity of the exoplanet host star $\iota$ Hor. To do so, we used the Ca \II\- H and K lines as a proxy for stellar activity. We analyzed
the line core fluxes of 987 CASLEO and HARPS spectra taken between
2002 and 2015. We detected a long-term activity cycle with a period of
4.57 years with a Monte Carlo FAP of 2 x 10$^{-5}$. \citet{2010ApJ...723L.213M} reported the
presence of a 1.6-yr chromospheric cycle for this star, which was
detected by using the Ca \II \ lines from SMART observations.
\citet{2013A&A...553L...6S} also found a 1.6-yr 
cycle in $\iota$ Hor in the X-ray emission, which is a direct indicator of
the coronal activity of the star. 

Considering that some stars show signatures of
  multiple cycles
  \citep[e.g.][]{1995ApJ...438..269B,2007AJ....133..862H,2009A&A...501..703O,2013ApJ...763L..26M,2016arXiv160600055R}
  and based in the results of \cite{2007ApJ...657..486B},
  \citet{2010ApJ...723L.213M} and \citet{2013A&A...553L...6S}
  suggested that $\iota$ Hor could belong to this type of stars. They
  estimated a longer cycle with a period around 6-yr, which  is
  apparently modulating the shorter one. In fact, from  Fig. 1 of
  \cite{2007ApJ...657..486B} it can be estimated a period for the
  long-term activity cycle (active branch) between 4 and 7-yr. The
  period we found in the present work fits this active branch within the
  statistical dispersion of the data in \citet{2007ApJ...657..486B}. 
 Furthermore, the stellar activity cycles are not exactly periodic
  \citep[e.g.][]{2009A&A...501..703O}. For instance, the Sun has an
  average cycle of almost exactly 11 years, however the length of the
  individual solar cycles fluctuates between 9 and 13 years with a
  standard deviation of about 1.16 years \citep[see details in][]{2010ASPC..428..307H}. In addition, it has a longer cycle with a period of 
80$\sim$90-year, called the Gleissberg cycle \citep[see e.g.,][]{2015LRSP...12....4H} and  quasi-biennial variations reported by \cite{2010ApJ...718L..19F}.

  An interesting aspect of our time series of Mount Wilson index
  of $\iota$ Hor is the ``flat behaviour'' observed between 
  2007 and 2012. In contrast, during that period,
  \citet{2010ApJ...723L.213M} and \citet{2013A&A...553L...6S} found
  the short cycle for $\iota$ Hor. To interpret these differences, we plotted the SMARTS averaged
  measurements together with CASLEO and HARPS data sets. On one hand,
  this figure shows in general good agreement between the different
  data sets. In fact, we can observe that our new measurements follow
  the expected increase of the new cycle. On the other hand, the
  analysis of the entire dataset confirm the
  1.6-year short activity cycle. These facts suggest that the
  differences are mainly produced by our sampling, and that frequent
  observations are necessary to better establish the
  presence of an inactive phase.

Our results suggest that both the shorter cycle (1.6-yr) and the
longer one (4.57-yr) are not regular, which is possibly related with
the high inclination of $\iota$ Hor (i $\approx$ 60$^{\circ}$) and the
antisymmetric distribution of active regions on the stellar surface,
as suggested by \citet{2013A&A...553L...6S}. 

\cite{2000A&A...353L..33K} detected a giant planet ($m\,sin\, i=2.26
M_{Jup}$) orbiting this star, with a period of ($320.1 \pm 2.1$)
days. There are several examples in the
literature where a long-period chromospheric signal mimick a planetary
signal \citep[e.g.][]{2010ApJ...725..875I,2011A&A...534A..30G,2013ApJ...774..147R,2014A&A...567A..48C}.
In order to detect a possible second planet in $\iota$ Hor, any possible periodic
signal should be disentangled to correctly identify its planetary
origin. In particular, \cite{2011A&A...528A...4B} excluded the
presence of a second low-mass planets with a period between 0.7 and
2.4 days in $\iota$ Hor. In our work we clearly
identified a possible $\sim$ 5-yr long-term cycle related to 
chromospheric activity, which could point toward the exclusion of a
second planet with a similar period.

\bibliographystyle{mn2e}
\bibliography{references}

\begin{thebibliography}{}

\bibitem[\protect\citeauthoryear{{Baliunas} \& {Soon}}{{Baliunas} \&
  {Soon}}{1995}]{1995ApJ...450..896B}
{Baliunas} S.,  {Soon} W.,  1995, \apj, 450, 896

\bibitem[\protect\citeauthoryear{{Baliunas},  \& {et al.}}{{Baliunas}
  et~al.}{1995}]{1995ApJ...438..269B}
{Baliunas} S.~L.,     {et al.} 1995, \apj, 438, 269

\bibitem[\protect\citeauthoryear{{B{\"o}hm-Vitense}}{{B{\"o}hm-Vitense}}{2007}]{2007ApJ...657..486B}
{B{\"o}hm-Vitense} E.,  2007, \apj, 657, 486

\bibitem[\protect\citeauthoryear{{Boisse}, {Bouchy}, {H{\'e}brard}, {Bonfils},
  {Santos} \& {Vauclair}}{{Boisse} et~al.}{2011}]{2011A&A...528A...4B}
{Boisse} I.,  {Bouchy} F.,  {H{\'e}brard} G.,  {Bonfils} X.,  {Santos} N.,
  {Vauclair} S.,  2011, \aap, 528, A4

\bibitem[\protect\citeauthoryear{{Bonfanti}, {Ortolani} \&
  {Nascimbeni}}{{Bonfanti} et~al.}{2016}]{2016A&A...585A...5B}
{Bonfanti} A.,  {Ortolani} S.,    {Nascimbeni} V.,  2016, \aap, 585, A5

\bibitem[\protect\citeauthoryear{{Buccino}, {D{\'{\i}}az}, {Luoni}, {Abrevaya}
  \& {Mauas}}{{Buccino} et~al.}{2011}]{2011AJ....141...34B}
{Buccino} A.~P.,  {D{\'{\i}}az} R.~F.,  {Luoni} M.~L.,  {Abrevaya} X.~C.,
  {Mauas} P.~J.~D.,  2011, \aj, 141, 34

\bibitem[\protect\citeauthoryear{{Buccino} \& {Mauas}}{{Buccino} \&
  {Mauas}}{2008}]{2008A&A...483..903B}
{Buccino} A.~P.,  {Mauas} P.~J.~D.,  2008, \aap, 483, 903

\bibitem[\protect\citeauthoryear{{Buccino} \& {Mauas}}{{Buccino} \&
  {Mauas}}{2009}]{2009A&A...495..287B}
{Buccino} A.~P.,  {Mauas} P.~J.~D.,  2009, \aap, 495, 287

\bibitem[\protect\citeauthoryear{{Buccino}, {Petrucci}, {Jofr{\'e}} \&
  {Mauas}}{{Buccino} et~al.}{2014}]{2014ApJ...781L...9B}
{Buccino} A.~P.,  {Petrucci} R.,  {Jofr{\'e}} E.,    {Mauas} P.~J.~D.,  2014,
  \apjl, 781, L9

\bibitem[\protect\citeauthoryear{{Carolo}, {Desidera}, {Gratton}, {Martinez
  Fiorenzano}, {Marzari}, {Endl}, {Mesa}, {Barbieri}, {Cecconi}, {Claudi},
  {Cosentino} \& {Scuderi}}{{Carolo} et~al.}{2014}]{2014A&A...567A..48C}
{Carolo} E.,  {Desidera} S.,  {Gratton} R.,  {Martinez Fiorenzano} A.~F.,
  {Marzari} F.,  {Endl} M.,  {Mesa} D.,  {Barbieri} M.,  {Cecconi} M.,
  {Claudi} R.~U.,  {Cosentino} R.,    {Scuderi} S.,  2014, \aap, 567, A48

\bibitem[\protect\citeauthoryear{{Cincunegui}, {D{\'{\i}}az} \&
  {Mauas}}{{Cincunegui} et~al.}{2007a}]{2007A&A...461.1107C}
{Cincunegui} C.,  {D{\'{\i}}az} R.~F.,    {Mauas} P.~J.~D.,  2007a, \aap, 461,
  1107

\bibitem[\protect\citeauthoryear{{Cincunegui} \& {Mauas}}{{Cincunegui} \&
  {Mauas}}{2004}]{2004A&A...414..699C}
{Cincunegui} C.,  {Mauas} P.~J.~D.,  2004, \aap, 414, 699

\bibitem[\protect\citeauthoryear{{Cincunegui}, {D{\'{\i}}az} \&
  {Mauas}}{{Cincunegui} et~al.}{2007b}]{2007A&A...469..309C}
{Cincunegui} C.~C.,  {D{\'{\i}}az} R.~F.,    {Mauas} P.~J.~D.,  2007b, \aap,
  469, 309

\bibitem[\protect\citeauthoryear{{Cuntz}, {Saar} \& {Musielak}}{{Cuntz}
  et~al.}{2000}]{2000ApJ...533L.151C}
{Cuntz} M.,  {Saar} S.~H.,    {Musielak} Z.~E.,  2000, \apjl, 533, L151

\bibitem[\protect\citeauthoryear{{D{\'{\i}}az}, {Cincunegui} \&
  {Mauas}}{{D{\'{\i}}az} et~al.}{2007}]{2007MNRAS.378.1007D}
{D{\'{\i}}az} R.~F.,  {Cincunegui} C.,    {Mauas} P.~J.~D.,  2007, \mnras, 378,
  1007

\bibitem[\protect\citeauthoryear{{D{\'{\i}}az}, {Gonz{\'a}lez}, {Cincunegui} \&
  {Mauas}}{{D{\'{\i}}az} et~al.}{2007}]{2007A&A...474..345D}
{D{\'{\i}}az} R.~F.,  {Gonz{\'a}lez} J.~F.,  {Cincunegui} C.,    {Mauas}
  P.~J.~D.,  2007, \aap, 474, 345

\bibitem[\protect\citeauthoryear{{Donati}, {Moutou}, {Far{\`e}s}, {Bohlender},
  {Catala}, {Deleuil}, {Shkolnik}, {Cameron}, {Jardine} \& {Walker}}{{Donati}
  et~al.}{2008}]{2008MNRAS.385.1179D}
{Donati} J.-F.,  {Moutou} C.,  {Far{\`e}s} R.,  {Bohlender} D.,  {Catala} C.,
  {Deleuil} M.,  {Shkolnik} E.,  {Cameron} A.~C.,  {Jardine} M.~M.,    {Walker}
  G.~A.~H.,  2008, \mnras, 385, 1179

\bibitem[\protect\citeauthoryear{{Fletcher}, {Broomhall}, {Salabert}, {Basu},
  {Chaplin}, {Elsworth}, {Garcia} \& {New}}{{Fletcher}
  et~al.}{2010}]{2010ApJ...718L..19F}
{Fletcher} S.~T.,  {Broomhall} A.-M.,  {Salabert} D.,  {Basu} S.,  {Chaplin}
  W.~J.,  {Elsworth} Y.,  {Garcia} R.~A.,    {New} R.,  2010, \apjl, 718, L19

\bibitem[\protect\citeauthoryear{{Flores}, {Gonz{\'a}lez}, {Jaque Arancibia},
  {Buccino} \& {Saffe}}{{Flores} et~al.}{2016}]{2016A&A...589A.135F}
{Flores} M.,  {Gonz{\'a}lez} J.~F.,  {Jaque Arancibia} M.,  {Buccino} A.,
  {Saffe} C.,  2016, \aap, 589, A135

\bibitem[\protect\citeauthoryear{{Gomes da Silva}, {Santos}, {Bonfils},
  {Delfosse}, {Forveille} \& {Udry}}{{Gomes da Silva}
  et~al.}{2011}]{2011A&A...534A..30G}
{Gomes da Silva} J.,  {Santos} N.~C.,  {Bonfils} X.,  {Delfosse} X.,
  {Forveille} T.,    {Udry} S.,  2011, \aap, 534, A30

\bibitem[\protect\citeauthoryear{{Hall}, {Lockwood} \& {Skiff}}{{Hall}
  et~al.}{2007}]{2007AJ....133..862H}
{Hall} J.~C.,  {Lockwood} G.~W.,    {Skiff} B.~A.,  2007, \aj, 133, 862

\bibitem[\protect\citeauthoryear{{Hathaway}}{{Hathaway}}{2010}]{2010ASPC..428..307H}
{Hathaway} D.~H.,  2010, in {ASP Conf. Ser. 428, SOHO 23: Understanding a
  Peculiar Solar Minimum} ed., {{Cranmer}, S.~R., {Hoeksema}, J.~T., {Kohl},
  J.~L.,} Vol.~428.
p.~307

\bibitem[\protect\citeauthoryear{{Hathaway}}{{Hathaway}}{2015}]{2015LRSP...12....4H}
{Hathaway} D.~H.,  2015, Living Reviews in Solar Physics, 12

\bibitem[\protect\citeauthoryear{{Horne} \& {Baliunas}}{{Horne} \&
  {Baliunas}}{1986}]{1986ApJ...302..757H}
{Horne} J.~H.,  {Baliunas} S.~L.,  1986, \apj, 302, 757

\bibitem[\protect\citeauthoryear{{Isaacson} \& {Fischer}}{{Isaacson} \&
  {Fischer}}{2010}]{2010ApJ...725..875I}
{Isaacson} H.,  {Fischer} D.,  2010, \apj, 725, 875

\bibitem[\protect\citeauthoryear{{Judge} \& {Saar}}{{Judge} \&
  {Saar}}{2007}]{2007ApJ...663..643J}
{Judge} P.~G.,  {Saar} S.~H.,  2007, \apj, 663, 643

\bibitem[\protect\citeauthoryear{{Krej{\v c}ov{\'a}} \& {Budaj}}{{Krej{\v
  c}ov{\'a}} \& {Budaj}}{2012}]{2012A&A...540A..82K}
{Krej{\v c}ov{\'a}} T.,  {Budaj} J.,  2012, \aap, 540, A82

\bibitem[\protect\citeauthoryear{{K{\"u}rster}, {Endl}, {Els}, {Hatzes},
  {Cochran}, {D{\"o}bereiner} \& {Dennerl}}{{K{\"u}rster}
  et~al.}{2000}]{2000A&A...353L..33K}
{K{\"u}rster} M.,  {Endl} M.,  {Els} S.,  {Hatzes} A.~P.,  {Cochran} W.~D.,
  {D{\"o}bereiner} S.,    {Dennerl} K.,  2000, \aap, 353, L33

\bibitem[\protect\citeauthoryear{{Lovis}, {Dumusque}, {Santos}, {Bouchy},
  {Mayor}, {Pepe}, {Queloz}, {S{\'e}gransan} \& {Udry}}{{Lovis}
  et~al.}{2011}]{2011arXiv1107.5325L}
{Lovis} C.,  {Dumusque} X.,  {Santos} N.~C.,  {Bouchy} F.,  {Mayor} M.,  {Pepe}
  F.,  {Queloz} D.,  {S{\'e}gransan} D.,    {Udry} S.,  2011, \arxiv

\bibitem[\protect\citeauthoryear{{Mengel}, {Fares}, {Marsden}, {Carter},
  {Jeffers}, {Petit}, {Donati}, {Folsom} \& {the BCool Collaboration}}{{Mengel}
  et~al.}{2016}]{2016MNRAS.459.4325M}
{Mengel} M.~W.,  {Fares} R.,  {Marsden} S.~C.,  {Carter} B.~D.,  {Jeffers}
  S.~V.,  {Petit} P.,  {Donati} J.-F.,  {Folsom} C.~P.,    {the BCool
  Collaboration} 2016, mnras, 459, 4325

\bibitem[\protect\citeauthoryear{{Metcalfe}, {Basu}, {Henry}, {Soderblom},
  {Judge}, {Kn{\"o}lker}, {Mathur} \& {Rempel}}{{Metcalfe}
  et~al.}{2010}]{2010ApJ...723L.213M}
{Metcalfe} T.~S.,  {Basu} S.,  {Henry} T.~J.,  {Soderblom} D.~R.,  {Judge}
  P.~G.,  {Kn{\"o}lker} M.,  {Mathur} S.,    {Rempel} M.,  2010, \apjl, 723,
  L213

\bibitem[\protect\citeauthoryear{{Metcalfe}, {Buccino}, {Brown}, {Mathur},
  {Soderblom}, {Henry}, {Mauas}, {Petrucci}, {Hall} \& {Basu}}{{Metcalfe}
  et~al.}{2013}]{2013ApJ...763L..26M}
{Metcalfe} T.~S.,  {Buccino} A.~P.,  {Brown} B.~P.,  {Mathur} S.,  {Soderblom}
  D.~R.,  {Henry} T.~J.,  {Mauas} P.~J.~D.,  {Petrucci} R.,  {Hall} J.~C.,
  {Basu} S.,  2013, \apjl, 763, L26

\bibitem[\protect\citeauthoryear{{Metcalfe}, {Judge}, {Basu}, {Henry},
  {Soderblom}, {Knoelker} \& {Rempel}}{{Metcalfe}
  et~al.}{2009}]{2009arXiv0909.5464M}
{Metcalfe} T.~S.,  {Judge} P.~G.,  {Basu} S.,  {Henry} T.~J.,  {Soderblom}
  D.~R.,  {Knoelker} M.,    {Rempel} M.,  2009, \ArXi

\bibitem[\protect\citeauthoryear{{Montes}, {L{\'o}pez-Santiago}, {G{\'a}lvez},
  {Fern{\'a}ndez-Figueroa}, {De Castro} \& {Cornide}}{{Montes}
  et~al.}{2001}]{2001MNRAS.328...45M}
{Montes} D.,  {L{\'o}pez-Santiago} J.,  {G{\'a}lvez} M.~C.,
  {Fern{\'a}ndez-Figueroa} M.~J.,  {De Castro} E.,    {Cornide} M.,  2001,
  \mnras, 328, 45

\bibitem[\protect\citeauthoryear{{Noyes}, {Hartmann}, {Baliunas}, {Duncan} \&
  {Vaughan}}{{Noyes} et~al.}{1984}]{1984ApJ...279..763N}
{Noyes} R.~W.,  {Hartmann} L.~W.,  {Baliunas} S.~L.,  {Duncan} D.~K.,
  {Vaughan} A.~H.,  1984, \apj, 279, 763

\bibitem[\protect\citeauthoryear{{Ol{\'a}h}, {Koll{\'a}th}, {Granzer},
  {Strassmeier}, {Lanza}, {J{\"a}rvinen}, {Korhonen}, {Baliunas}, {Soon},
  {Messina} \& {Cutispoto}}{{Ol{\'a}h} et~al.}{2009}]{2009A&A...501..703O}
{Ol{\'a}h} K.,  {Koll{\'a}th} Z.,  {Granzer} T.,  {Strassmeier} K.~G.,  {Lanza}
  A.~F.,  {J{\"a}rvinen} S.,  {Korhonen} H.,  {Baliunas} S.~L.,  {Soon} W.,
  {Messina} S.,    {Cutispoto} G.,  2009, aap, 501, 703

\bibitem[\protect\citeauthoryear{{Pace}}{{Pace}}{2013}]{2013A&A...551L...8P}
{Pace} G.,  2013, \aap, 551, L8

\bibitem[\protect\citeauthoryear{{Parker}}{{Parker}}{1955}]{1955ApJ...122..293P}
{Parker} E.~N.,  1955, \apj, 122, 293

\bibitem[\protect\citeauthoryear{{Poppenhaeger} \& {Wolk}}{{Poppenhaeger} \&
  {Wolk}}{2014}]{2014A&A...565L...1P}
{Poppenhaeger} K.,  {Wolk} S.~J.,  2014, \aap, 565, L1

\bibitem[\protect\citeauthoryear{{Radick}, {Lockwood}, {Skiff} \&
  {Thompson}}{{Radick} et~al.}{1995}]{1995ApJ...452..332R}
{Radick} R.~R.,  {Lockwood} G.~W.,  {Skiff} B.~A.,    {Thompson} D.~T.,  1995,
  \apj, 452, 332

\bibitem[\protect\citeauthoryear{{Rebull},  \& {et al.}}{{Rebull}
  et~al.}{2016}]{2016arXiv160600055R}
{Rebull} L.~M.,     {et al.} 2016, \arxivs

\bibitem[\protect\citeauthoryear{{Robertson}, {Endl}, {Cochran}, {MacQueen} \&
  {Boss}}{{Robertson} et~al.}{2013}]{2013ApJ...774..147R}
{Robertson} P.,  {Endl} M.,  {Cochran} W.~D.,  {MacQueen} P.~J.,    {Boss}
  A.~P.,  2013, \apj, 774, 147

\bibitem[\protect\citeauthoryear{{Robinson} \& {Durney}}{{Robinson} \&
  {Durney}}{1982}]{1982AA...108..322}
{Robinson} R.~D.,  {Durney} B.~R.,  1982, \aap, 108, 322

\bibitem[\protect\citeauthoryear{{Saar} \& {Brandenburg}}{{Saar} \&
  {Brandenburg}}{1999}]{1999ApJ...524..295S}
{Saar} S.~H.,  {Brandenburg} A.,  1999, \apj, 524, 295

\bibitem[\protect\citeauthoryear{{Sanz-Forcada}, {Stelzer} \&
  {Metcalfe}}{{Sanz-Forcada} et~al.}{2013}]{2013A&A...553L...6S}
{Sanz-Forcada} J.,  {Stelzer} B.,    {Metcalfe} T.~S.,  2013, \aap, 553, L6

\bibitem[\protect\citeauthoryear{{Shkolnik}, {Bohlender}, {Walker} \& {Collier
  Cameron}}{{Shkolnik} et~al.}{2008}]{2008ApJ...676..628S}
{Shkolnik} E.,  {Bohlender} D.~A.,  {Walker} G.~A.~H.,    {Collier Cameron} A.,
   2008, \apj, 676, 628

\bibitem[\protect\citeauthoryear{{Vauclair}, {Laymand}, {Bouchy}, {Vauclair},
  {Hui Bon Hoa}, {Charpinet} \& {Bazot}}{{Vauclair}
  et~al.}{2008}]{2008A&A...482L...5V}
{Vauclair} S.,  {Laymand} M.,  {Bouchy} F.,  {Vauclair} G.,  {Hui Bon Hoa} A.,
  {Charpinet} S.,    {Bazot} M.,  2008, \aap, 482, L5

\bibitem[\protect\citeauthoryear{{Vaughan}, {Preston} \& {Wilson}}{{Vaughan}
  et~al.}{1978}]{1978PASP...90..267V}
{Vaughan} A.~H.,  {Preston} G.~W.,    {Wilson} O.~C.,  1978, \mnras, 90, 267

\bibitem[\protect\citeauthoryear{{Vieytes}, {Mauas} \& {Cincunegui}}{{Vieytes}
  et~al.}{2005}]{2005A&A...441..701V}
{Vieytes} M.,  {Mauas} P.,    {Cincunegui} C.,  2005, \aap, 441, 701

\bibitem[\protect\citeauthoryear{{Vieytes}, {Mauas} \& {D{\'{\i}}az}}{{Vieytes}
  et~al.}{2009}]{2009MNRAS.398.1495V}
{Vieytes} M.~C.,  {Mauas} P.~J.~D.,    {D{\'{\i}}az} R.~F.,  2009, \mnras, 398,
  1495

\bibitem[\protect\citeauthoryear{{Walker}, {Croll}, {Matthews}, {Kuschnig},
  {Huber}, {Weiss}, {Shkolnik}, {Rucinski}, {Guenther}, {Moffat} \&
  {Sasselov}}{{Walker} et~al.}{2008}]{2008A&A...482..691W}
{Walker} G.~A.~H.,  {Croll} B.,  {Matthews} J.~M.,  {Kuschnig} R.,  {Huber} D.,
   {Weiss} W.~W.,  {Shkolnik} E.,  {Rucinski} S.~M.,  {Guenther} D.~B.,
  {Moffat} A.~F.~J.,    {Sasselov} D.,  2008, \aap, 482, 691

\bibitem[\protect\citeauthoryear{{White}, {Skumanich}, {Lean}, {Livingston} \&
  {Keil}}{{White} et~al.}{1992}]{1992PASP..104.1139W}
{White} O.~R.,  {Skumanich} A.,  {Lean} J.,  {Livingston} W.~C.,    {Keil}
  S.~L.,  1992, \pasp, 104, 1139

\bibitem[\protect\citeauthoryear{{Wilson}}{{Wilson}}{1978}]{1978ApJ...226..379W}
{Wilson} O.~C.,  1978, \apj, 226, 379

\bibitem[\protect\citeauthoryear{{Wright}}{{Wright}}{2004}]{2004AJ....128.1273W}
{Wright} J.~T.,  2004, \aj, 128, 1273

\end{thebibliography}

\appendix

\end{document}